\documentclass[12pt]{article}
\pdfoutput=1
\usepackage{mathrsfs}
\usepackage{natbib,graphicx,setspace,lscape,longtable,xr}
\usepackage{mathrsfs,amsmath,amsthm,amssymb,color}
\usepackage{natbib,epsfig,graphicx,pdfpages}
\usepackage{rotating}
\usepackage{float}
\usepackage{bm}
\usepackage{ulem}
\usepackage{CJK}
\usepackage{ragged2e}
\usepackage{setspace}
\usepackage{threeparttable}
\usepackage{multirow}
\usepackage{enumitem}
\usepackage{hyperref}
\usepackage{etoolbox}
\usepackage{subfig}
\usepackage[ruled]{algorithm2e}
\usepackage{adjustbox}
\usepackage{blindtext}

\bibpunct{(}{)}{;}{a}{,}{,}

\newtheorem{theorem}{Theorem}
\newtheorem{lemma}{Lemma}
\newtheorem{proposition}{Proposition}
\newcommand{\mylabel}[2]{#2\def\@currentlabel{#2}\label{#1}}
\newcommand{\csection}[1]
{\begin{center}
		\stepcounter{section}
		{\bf\large\arabic{section}. #1}
	\end{center}
}

\newcommand{\csubsection}[1]{
	\begin{center}
		\stepcounter{subsection}
		{\it\arabic{section}.\arabic{subsection}. #1}
	\end{center}
}

\newcommand{\scsubsection}[1]{
	\begin{center}
		\stepcounter{subsection}
		{\it #1}
	\end{center}
}

\def\beq{\begin{equation}}
	\def\eeq{\end{equation}}
\def\beqr{\begin{eqnarray}}
	\def\eeqr{\end{eqnarray}}
\def\beqrs{\begin{eqnarray*}}
	\def\eeqrs{\end{eqnarray*}}
\def\bet{\begin{theorem}}
	\def\eet{\end{theorem}}
\def\bel{\begin{lemma}}
	\def\eel{\end{lemma}}
\def\bep{\begin{proposition}}
	\def\eep{\end{proposition}}
\def\bg{\begin{figure}[tbph]\begin{center}}
		\def\eg{\end{center}\end{figure}}

\def\bc{\begin{center}}
	\def\ec{\end{center}}

\def\wt{\widetilde}

\def\wh{\widehat}

\def\mR{\mathbb{R}}
\def\mL{\mathcal L}

\def\mS{\mathcal S}

\def\mW{\mathbb{W}}

\def\mX{\mathbb{X}}

\def\mY{\mathbb{Y}}

\def\argmax{\mbox{argmax}}

\renewcommand{\arraystretch}{1.3}

\usepackage[a4paper,bindingoffset=0.2in,%
left=1in,right=1in,top=1in,bottom=1in,%
footskip=.25in]{geometry}

\def\boxit#1{\vbox{\hrule\hbox{\vrule\kern6pt\vbox{\kern6pt#1\kern6pt}\kern6pt\vrule}\hrule}}

\numberwithin{equation}{section}

\begin{document}
\begin{CJK}{GBK}{song}
\begin{center}
{\bf\Large Distributed Logistic Regression for Massive Data with Rare Events}\\
\bigskip

Xuetong Li$^{1}$, Xuening Zhu$^{2}$, and Hansheng Wang$^1$

{\it\small

$^1$ Guanghua School of Management, Peking University, Beijing, China;\\
$^2$ School of Data Science, Fudan University, Shanghai, China.
}
\end{center}

\begin{singlespace}
\begin{abstract}
Large-scale rare events data are commonly encountered in practice.
To tackle the massive rare events data, we propose a novel distributed estimation method for logistic regression in a distributed system.
For a distributed framework, we face the following two challenges.
The first challenge is how to distribute the data.
In this regard, two different distribution strategies (i.e., the RANDOM strategy and the COPY strategy) are investigated.
The second challenge is how to select an appropriate type of objective function so that the best asymptotic efficiency can be achieved.
Then, {\color{black} the under-sampled (US) and inverse probability weighted (IPW) types of objective functions} are considered.
Our results suggest that the COPY strategy together with the {\color{black} IPW objective function} is the best solution for distributed logistic regression with rare events.
The finite sample performance of the distributed methods is demonstrated by simulation studies and a real-world Sweden Traffic Sign dataset.

\end{abstract}


\end{singlespace}

\newpage

\csection{INTRODUCTION}

\begin{figure}[htb]
\centering
\includegraphics[width=0.6 \textwidth]{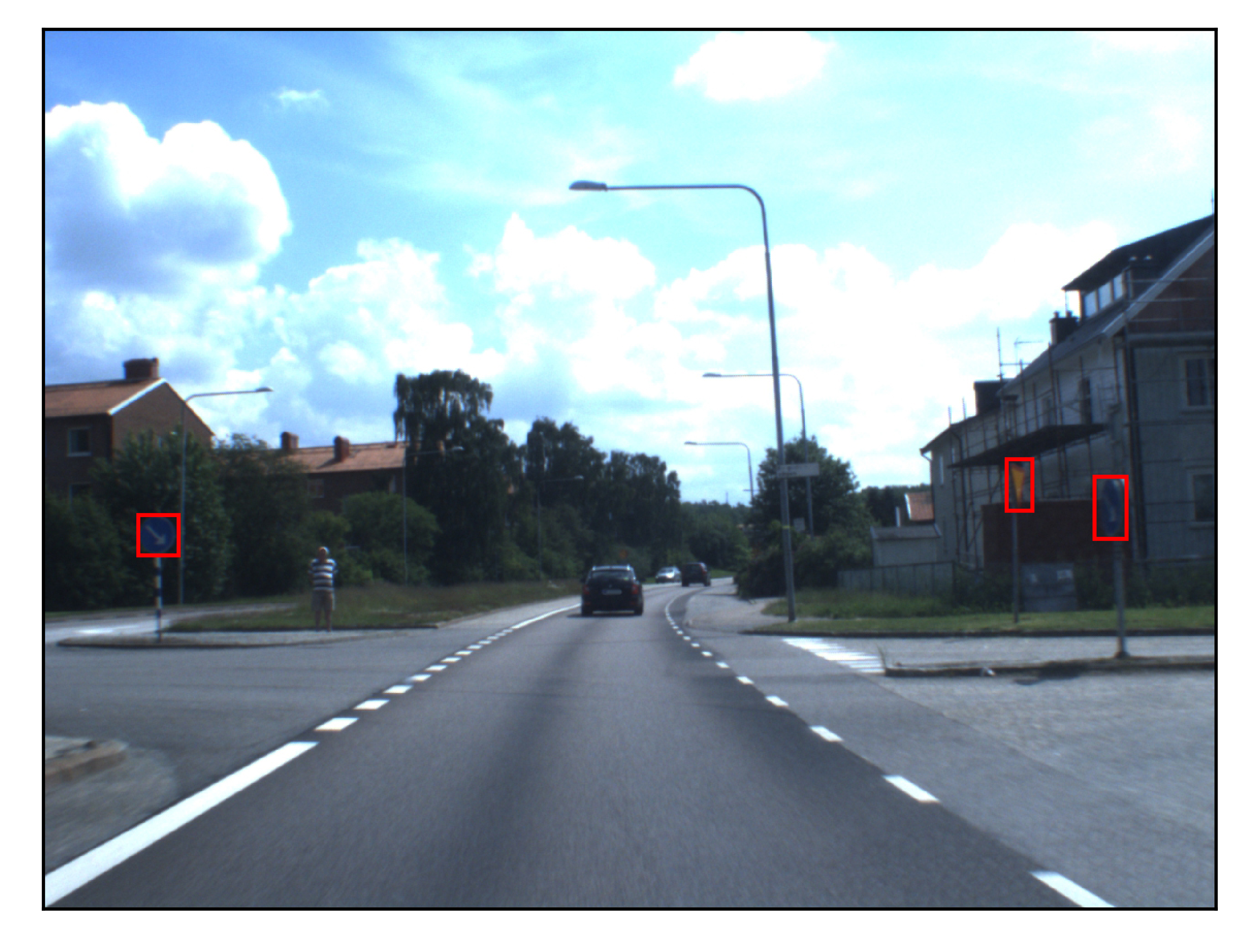}\hfill
\caption{
	An example in the Sweden Traffic Sign dataset for traffic sign detection.
	The original image is of size $960 \times 1,280 \times 3$. Each red bounding box is used to annotate a local region containing a traffic sign.
	The bounding box of a small object treated as a positive instance only covers less than 1\% of the original image.}
\label{f: bbox}
\end{figure}

Massive data with rare events in binary regression are commonly encountered in scientific fields and applications.
Conceptually, rare events data, also called imbalanced data, refer to the number of instances in the positive class being much smaller than that in the negative class.
For example, in online search or recommendation systems, billions of impressions can be generated each day.
If we treat each impression as one sample, then the probability for one impression to generate a click is very small.
Thus, clicks could be treated as rare events \citep{japkowicz2000learning, mcmahan2013ad, chen2016deep, huang2020embedding}.
As another example in political science, the occurrence of wars, vetos, coups and the decisions of citizens to run for office have been modeled as rare events \citep{king2001logistic, owen2007infinitely,neunhoeffer2019cross}.
Our last example is small object detection in a high resolution image; see Figure \ref{f: bbox}.
Suppose we treat each pixel as a sample and whether it is covered by a bounding box as corresponding response.
Then, the bounding box of a small object treated as a positive instance only covers less than 1\% of the original image \citep{zhu2016traffic, zhao2019object, chen2020survey}.
Other important rare events data examples include fraud
detection \citep{bolton2002statistical, hassan2016modeling}, drug discovery \citep{zhu2006lago,korkmaz2020deep} and
rare disease diagnosis \citep{zhao2018framework,zhuang2019care}.
For a comprehensive summary, we refer to \cite{sun2009classification}, \cite{haixiang2017learning} and \cite{kaur2019systematic}.

A common approach to tackle imbalanced data is to balance it by under-sampling the negative class \citep{drummond2003c4, liu2008exploratory,nguyen2012comparative} or oversampling the positive class \citep{chawla2002smote,han2005borderline,mathew2017classification}.
Most existing literature focuses on practical algorithms and methodologies for classification with few statistical theory guarantees.
They design sampling strategies or ensemble learning methods to improve classification accuracy \citep{krawczyk2016learning}.
For example,
\cite{estabrooks2004multiple} empirically investigated the effective combination of different resampling paradigms to improve classification accuracy. \cite{sun2007cost} adapted the AdaBoost algorithm for advancing the classification of imbalanced data.
\cite{king2001logistic} considered logistic regression in rare events data and focused on correcting the biases when estimating regression coefficients and probabilities.
\cite{fithian2014local} used the special structure of logistic regression models to design a novel local case-control sampling method.
However, these theoretical studies are based on the regular assumption that the probability of event occurring is fixed.
This might not be the best way to describe rare events mathematically,
because this assumption implies that the number of rare events should diverge to infinity at the same rate as the total sample size diverges towards infinity.
Instead, for rare events, it is more appropriate to assume that the positive class rate should decay towards zero as the total sample size increases.

In this regard, \cite{wang2020logistic} developed a novel theoretical framework and the resulting estimators' statistical properties were investigated accordingly.
Under his novel theoretical framework, he showed that the convergence rate of the global maximum likelihood estimator (GMLE) is mainly determined by the number of positive instances instead of the total sample size.
As a consequence, the convergence rate of the GMLE should be considerably slower than that of the usual cases. 
Additionally, \cite{wang2020logistic} surprisingly found that both under-sampling and over-sampling methods would cause unnecessary statistical efficiency loss in parameter estimation.
{\color{black} Then, how to develop new estimation methods so that a statistically efficient estimator can be obtained becomes a problem of great importance.
It is remarkable that we call an estimator to be statistically efficient, if it achieves the same asymptotic distribution as the GMLE, throughout the rest of this article.

It is worth mentioning that we are not among the first group of researchers studying the problem of logistic regression for massive data.
Significant progresses have been made in the past literature.
One possible solution is subsampling. 
For example, \cite{wang2018optimal} developed a subsampling method, which is motivated by the A-optimality criterion of \cite{kiefer1959optimum}. 
\cite{wang2019more} further proposed more efficient estimators based on subsamples with the optimal subsampling probabilities.
{\color{black} 
A general model with imbalanced binary response is studied by \cite{wang2021nonuniform} recently.
}
Another possible solution is distributed computing, if a parallel computing system can be used.
For example, \cite{du2018privacy} proposed differentially private approaches to collaboratively and accurately train a logistic regression model among multiple parties.
\cite{shi2019distributed} studied the distributed logistic regression based on the classical ADMM algorithm \citep{boyd2011distributed}.
\cite{zuo2021optimal} proposed a distributed subsampling procedure to approximate the maximum likelihood estimator. 
}
A cost-sensitive algorithm was developed by \cite{wang2016fast} for the linear SVM problem.
Despite the usefulness of the above methods,
very few attempts have been made for distributed classification problems with rare events data and rigorous asymptotic theory.
Without a solid theoretical guidance, we are not able to deliver a statistically efficient estimator in this regard. 
This motivates us to develop a novel distributed logistic regression method with solid statistical theory support for massive rare events data.

It is noteworthy that developing a distributed estimation method for logistic regression with rare events is not straightforward.
We face at least the following two challenging problems.
The first problem is data distribution on local computers in a distributed system.
Because the total number of positive instances is much smaller than the total sample size, the traditional pure random data distribution strategy might not be the best choice in some cases.
For example, if the number of instances assigned to the local machine is very small, this traditional strategy leads to even smaller positive instances for each distributed computer node.
This process makes the local estimates obtained from each local computer statistically inaccurate,
which in turn makes the finally combined estimator statistically inefficient.
In fact, a potentially better choice is to copy all the positive instances to each local computer and then the negative instances should be distributed to local computers as randomly as possible.
For convenience, we refer to the traditional data distribution strategy as a fully RANDOM strategy and this new strategy as a COPY strategy.
Then, investigating the statistical properties of the estimators
under both RANDOM and COPY strategies becomes a problem of great interest.

The second problem is the choice of objective function.
If the COPY strategy is adopted, the positive and negative instances become much more balanced on each local computer, which makes the statistical estimation easier.
However, the side effect is that the local objective function is no longer unbiased for the global log-likelihood function.
Thus, the resulting estimator is statistically inefficient,
even though the resulting estimator remains to be asymptotically normal.
This is an interesting finding of \cite{wang2020logistic}.
For convenience, we refer to this estimator computed on each local computer as an under-sampled estimator.
To solve this problem, a new-type objective function is proposed on each local computer, which should be unbiased for the global one.
This naturally leads to an inverse probability weighted estimator \citep{fithian2014local,wang2020logistic}.
Subsequently, we consider obtaining a distributed logistic regression estimator.
A simple and common approach is to take the average of estimators produced by local computers.
This approach is referred to the one-shot (OS) method in the literature \citep{zhang2013communication,rosenblatt2016optimality,chang2017divide}.
We use the OS method to combine the local IPW estimators to yield the final estimator, which is referred to as the IPW estimator.

To summarize, we aim to make the following important contributions to the existing literature.
{\color{black} First, we theoretically prove that the traditional RANDOM distributed framework cannot perform efficiently with rare events data due to its unignorable random bias term in many cases.} 
Second, a COPY strategy is proposed and rigorously investigated.
The US type of local objective function is used to construct a US estimator.
We find that the US estimator has a lower bias but unsatisfactory statistical efficiency if the number of negative instances on each computer node is not enough.
Lastly, we find that the IPW estimator is statistically more efficient than the US estimator and has the same asymptotic behavior as the GMLE.
Theoretical findings are further verified by extensive numerical studies.

The remainder of this paper is organized as follows.
Section 2 introduces the model setting and three important benchmark estimation methods according to \cite{wang2020logistic}.
Section 3 presents three distributed estimation methods and their asymptotic theory.
Numerical studies are given in Section 4.
An application to the Sweden traffic Sign Data is illustrated here using these three distributed methods.
The article concludes with a brief discussion in Section 5.
All technical details are delegated to the appendix.

\csection{LOGISTIC REGRESSION WITH RARE EVENTS DATA}

\csubsection{Model Setup}

Suppose there are $N$ observations in total, which are indexed by $1 \le i \le N$.
The $i$th observation is denoted as $\big(X_i, Y_i\big)$, where $X_i \in \mR^{p}$ is a $p$-dimensional covariate and $Y_i \in\{0,1\}$ is the binary response.
Assume $\big(X_i, Y_i\big)$ is independently generated for $1\le i\le N$ and denote the full data by $\mS_F=\big\{\big(X_{i}, Y_i\big): 1\le i\le N \big\}$.
Let $N_1=\sum_{i=1}^{N} Y_i$ be the number of positive instances, and $N_0=N-N_1$ be the number of negative instances.
To model their regression relationship, the following logistic regression model is considered
\beqr
\label{eq: logistic}
P\big(Y_i=1 \mid X_i\big) =p_i(\alpha, \beta)= \frac{e^{\alpha+X_i^\top \beta}}{1+e^{\alpha+X_i^\top \beta}},
\eeqr
where $\alpha \in \mR$ is the intercept and $\beta \in \mR^{p}$ is the slope parameter.
Define $\theta=(\alpha, \beta^{\top})^{\top} \in \mR^{p+1}$ as the full parameter vector with true value given by $\theta^* = (\alpha^*, \beta^{* \top})^{\top}$.
As $N$ diverges to infinity, if $\theta^*$ does not change, the number of positive instances would diverge at a rate of $O_p(N)$.
Following \cite{shao2003mathematical}, we define $O_p(\cdot)$ as follows.
Let $\{A_i\}$ and $\{B_i\}$ with $1\le i\le N$ be two random variable sequences. 
We then say $A_i=O_p(B_i)$ if and only if for any $\varepsilon > 0$ there is a constant $C_{\varepsilon} > 0$, such that $\sup_i P(\|A_i\| \ge C_{\varepsilon}\|B_i\|) < \varepsilon$.

Under the classical logistic regression model setting (\ref{eq: logistic}), existing theory shows that the maximum likelihood estimator (MLE) based on the full data $\mS_F$ converges at a rate of $O_p(N^{-1/2})$ \citep{nelder1972generalized}.
As convincingly argued by \cite{wang2020logistic}, this might not be the best choice for modeling rare events data.
For rare events data, the percentage of positive instances is extremely small.
Statistically, it is more appropriate to specify the positive response rate to converge towards 0 as the total sample size increases towards infinity.
Meanwhile, we wish the covariate effect (as measured by $\beta^*$) remains constant since the value of $X$ is unknown.
Otherwise, it cannot be accurately estimated statistically.
Consequently, this suggests that we should replace the intercept parameter $\alpha^*$ by $\alpha_{N}^*$, which should diverge towards negative infinity as $N \to
\infty$.
Specifically, we should have $\alpha_{N}^*\to-\infty $ at an appropriate divergence rate as $N \to \infty$.
However, what is a reasonable divergence rate requires more careful investigation.
Under this assumption, we should have $P\big(Y_i=1 \mid X_i\big) \approx e^{\alpha_{N}^*+X_i^\top \beta^*}$ as $N\to \infty$. We then have $E(N_1) \approx Ne^{\alpha_{N}^*} E\big(e^{X_i^\top \beta^*}\big)$. Even though the positive response rate (i.e., $N_1/N$) should converge toward 0 as $N$ goes to infinity, we still expect that the total number of positive instances (i.e., $N_1$) should diverge to infinity. Otherwise, we cannot estimate the parameters of interest consistently.
This suggests that we should have
\beqr
\label{eq: condition}
\alpha_{N}^*\to-\infty \quad \text{and} \quad  \alpha_{N}^*+\log N\to\infty,
\eeqr
when $N\to \infty$.
This becomes the most important technical assumption for the proposed theoretical framework \citep{wang2020logistic}.

\csubsection{Related Methods}

In this subsection, we demonstrate a number of important benchmark estimation methods according to \cite{wang2020logistic}.
Specifically, we introduce the global maximum likelihood estimation, under-sampled estimation, and inverse probability weighted likelihood estimation, respectively.

\scsubsection{Global Maximum Likelihood Estimation}

We start with the global maximum likelihood estimation method using the full data. The log-likelihood function based on the full data $\mS_F$ is given as follows:
\beqr
\label{eq: global likelihood}
\mL \big(\theta\big)=\sum_{i=1}^{N}\Big\{Y_i \log p_i\big(\alpha_N,\beta\big) + \big(1-Y_i\big)\log \big(1-p_i(\alpha_N,\beta) \big)\Big\},
\eeqr
where $p_i(\alpha_N, \beta) = e^{\alpha_N+X_i^\top \beta}/(1+e^{\alpha_N+X_i^\top \beta})$.
Then we could obtain the GMLE as $\wh\theta_{\rm GMLE} = \argmax_{\theta} \mL (\theta)$.
According to Theorem 1 in \cite{wang2020logistic}, the GMLE $\wh\theta_{\rm GMLE}$ should be $\sqrt{Ne^{\alpha_N^*}}-$consistent and asymptotically normal under appropriate conditions.
This result suggests that the convergence rate of the GMLE is fully determined by the number of positive instances, which implies that the help provided by an extra large amount of the negative instances should be limited.
This result is particularly true when the total number of negative instances is too large to be easily managed on one computer.

Nevertheless, we should remark that this never implies that a large number of negative instances is totally useless for efficiency improvement.
Extensive theoretical and numerical experiences suggest that the statistical efficiency of various benchmark estimators can be improved by a more efficient use of negative instances, even though the convergence rate remains unchanged.
However, for many practical datasets with rare events, the total number of negative instances is often too large to be easily managed on one computer.
In this case, how to utilize negative instances more efficiently for better estimation efficiency becomes a problem of great interest.

\scsubsection{Under-Sampled Estimation}

In practice, researchers often seek to include all the positive instances for statistical analysis, because they are rare and thus valuable \citep{drummond2003c4, liu2008exploratory,nguyen2012comparative}.
Next, the same (or comparable) number of negative instances are randomly selected so that a more balanced subsample can be constructed.
Subsequently, interested parameters can be estimated based on this more balanced subsample.
For convenience, we refer to this common practice as an under-sampled method \citep{drummond2003c4, liu2008exploratory, nguyen2012comparative,wang2020logistic}.
By doing so, the estimation problem becomes computationally feasible.
Theoretically, this problem can be formulated as follows.
Let $a_i$ be a binary indicator with $P(a_i=1)= \pi$, which is independently generated for each $i$.
Here, $a_i=1$ suggests that $i$th instance is sampled and
$\pi$ is the probability for sampling.
Accordingly, the US objective function \eqref{eq: logistic} becomes
\beqr
\label{eq: unweighted likelihood}
\mL_{\rm US}\big(\theta\big)=\sum_{i=1}^{N}\Big\{Y_i \log p_i\big(\alpha_N,\beta\big) + \big(1-Y_i\big)a_i\log \big(1-p_i(\alpha_N,\beta) \big)\Big\}.
\eeqr
{\color{black} 
For convenience, we call it a US objective function.}
Then, we obtain a US estimator as $\wh\theta_{\rm US} = \argmax_{\theta} \mL_{\rm US} (\theta)$.
However, \cite{wang2020logistic} finds that $\wh\theta_{\rm US}$ is a biased estimator for $\theta^*$.
Thus, the debiased US estimator is further obtained as $\wt\theta_{\rm US} = \wh\theta_{\rm US} + (\log \pi,0,\cdots,0)^\top$.

Comparing \eqref{eq: unweighted likelihood} with \eqref{eq: global likelihood}, we find the only difference is the treatment of the negative instances.
Considering \eqref{eq: global likelihood}, all the instances are used regardless of positives or negatives.
However, considering \eqref{eq: unweighted likelihood}, we use all positive instances, and negative instances are included only if the corresponding binary indicator $a_i=1$.
By doing so, we have all positive instances included and only a much smaller number of negative instances are used.
One can verify easily that this formulation is mathematically equivalent to that of \cite{wang2020logistic}.
The careful theoretical analysis of \cite{wang2020logistic} suggests that such an estimator remains to be $\sqrt{Ne^{\alpha_N^*}}$-consistent
and is asymptotically normal.
However, as shown in Theorem 3 by \cite{wang2020logistic},
the US estimator cannot obtain the same efficiency as the GMLE if the ratio of positive instances to the negative instances does not converge to zero.

\scsubsection{Inverse Probability Weighted Estimation}

The key reason for the statistical inefficiency of the US estimator is {\color{black} the objective function in (\ref{eq: unweighted likelihood}) }.
By under-sampling, the resulting {\color{black} objective function} has been materially changed.
A direct consequence is that it is no longer an unbiased estimator for the global log-likelihood function.
That leads to the inefficiency for the US estimator.
To fix this problem, one possible solution is to find an unbiased estimator for the global log-likelihood function. 
This leads to the following {\color{black} objective function} for inverse probability weighted estimation \citep{king2001logistic,fithian2014local,wang2020logistic}
\beqr
\label{eq: weighted likelihood}
\mL_{\rm IPW}\big(\theta\big) = \sum_{i=1}^{N} \Big\{Y_i \log p_i\big(\alpha_N,\beta\big) + \big(1-Y_i\big) a_i \log \big(1-p_i(\alpha_N,\beta) \big) / \pi \Big\}.
\eeqr
One can easily verify that $E \{\mL_{\rm IPW}(\theta)| \mS_F\}=\mL(\theta)$, which suggests that $\mL_{\rm IPW}\big(\theta\big)$ is an unbiased estimator for the global log-likelihood function.
By optimizing the above {\color{black} objective function}, an IPW estimator can be obtained as $\wh\theta_{\rm IPW} = \argmax_{\theta}\mL_{\rm IPW}(\theta)$.
{\color{black} \cite{wang2020logistic} demonstrated that the IPW-type estimator has the same convergence rate $O_p\big(1/\sqrt{Ne^{\alpha_N^*}}\big)$ as that of $\wh\theta_{\rm GMLE}$ but remains to be statistically inefficient.
Recall that we define in this work an estimator to be statistically efficient if it shares the same asymptotic distribution as the GMLE.}

The suboptimal efficiency of both the US and IPW estimators is understandable because both methods include only a very small fraction of the negative instances for estimation.
Then, there should exist a good possibility to use a larger number of negative instances (but not as large as the full set of negative class) for better statistical efficiency.
This seems to be a particularly promising direction if a powerful distributed computing system is available.
With the help of a distributed system, we should be able to compute various local estimators (e.g., the US and IPW estimators) multiple times.
They can then be aggregated together to form a more powerful estimator.
However, what type of local estimators should be computed and how they should be assembled so that the final estimator can be as efficient as the GMLE are problems of great interest.
We thus aim to systematically investigate these interesting problems in the next sections.

\csection{DISTRIBUTED LOGISTIC REGRESSION}

\csubsection{Distributed MLE with Random Strategy}

\begin{figure}[htbp]
\centering
\includegraphics[width=0.6\textwidth]{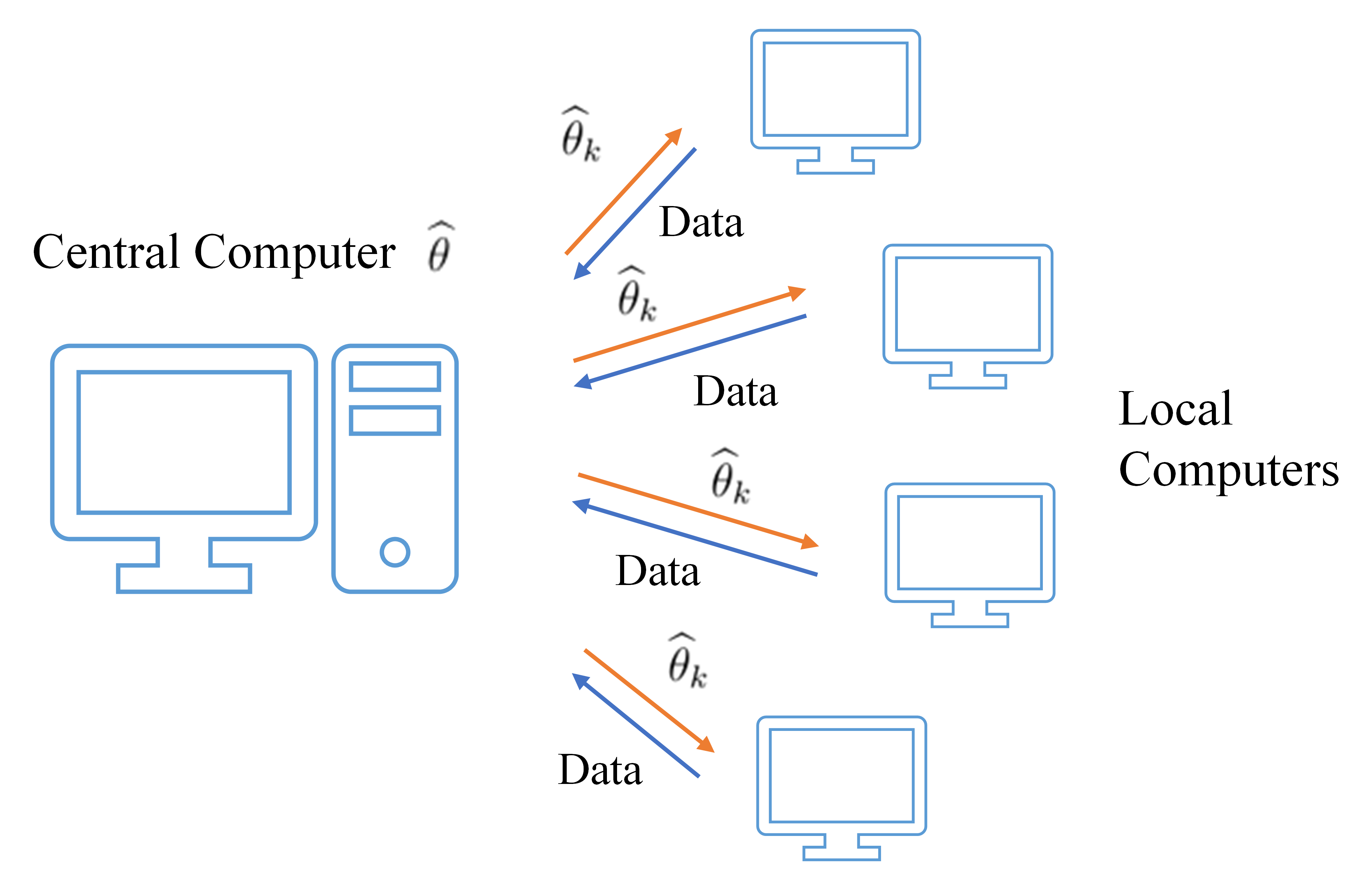}\hfill
\caption{Illustration of the distributed system.
}
\label{f: distributed system}
\end{figure}

We start with the simplest distributed estimator, that is the distributed maximum likelihood estimator obtained under the RANDOM strategy. For convenience, we refer to this as RMLE.
Assume there exists a distributed computation system with a total of $K$ local computers and one central computer.
A typical architecture of a distributed system is shown in Figure \ref{f: distributed system}.
The local computers are indexed by $1\le k\le K$.
Then, the RMLE method randomly distributes the full data $\mS_F$ to each local computer with approximately equal sizes.
Denote $\mS_F = \mS_+ \cup \mS_-$, where $\mS_+=\{i:Y_i=1\}$ represents the set of all the positive instances, and $\mS_-=\{i:Y_i=0\}$ represents the set of all negative instances.
Specifically, let $\mS_k^R$ be the sample randomly distributed to the $k$th local computer with $\mS_k^R = \mS_{k+}^R\cup \mS_{k-}^{R}$,
where $\mS_{k+}^R = \{i:i\in \mS_k^R, Y_i = 1\}$ and $\mS_{k-}^R
= \{i:i\in \mS_k^R, Y_i = 0\}$ refer to the set of positive and negative instances on the $k$th local computer, respectively.
For convenience, denote $n_k=|\mS_k^R|$.
In addition, let $n_{1k}^R = |\mS_{k+}^R|$
and $n_{0k}^R = |\mS_{k-}^R|$.
Mathematically, denote $a_i^{(k)}=1$ if the $i$th observation is randomly distributed to $k$th local computer.
We then have $\sum_{k=1}^K a_i^{(k)}=1$ for every $i$, $n_k = \sum_{i=1}^N a_i^{(k)}$.
We also define $n = E(n_k)= N/K$.
Additionally, we have $n_{1k} = \sum_{i=1}^N a_i^{(k)}Y_i$ and $n_{0k} = \sum_{i=1}^N a_i^{(k)}(1-Y_i)$.

As one can see, by the RANDOM strategy, both the positive and negative instances are randomly distributed to each local computer. As a consequence, their relative percentages remain approximately the same as the full data size. That is
$n_{1k}/n_k \approx N_1/N$.
The merit of this method is that the data distribution on each local computer remains the same as the full data.
However, the drawback is that the positive instances allocated to each local computer become even smaller. This might turn into statistical inefficiency for the resulting estimator.
Specifically, for each local computer, define
\beqr
\mL_{{\rm R},k} \big(\theta\big) = \sum_{i=1}^{N} a_i^{(k)} \Big\{Y_i \log p_i\big(\alpha_N,\beta\big) + \big(1-Y_i\big)\log \big(1-p_i(\alpha_N,\beta) \big)\Big\}
\eeqr
as a local log-likelihood function with $P(a_i^{(k)}=1)=1/K$.
Then a local MLE is computed as $\wh\theta_{{\rm RMLE},k}=\argmax_{\theta}\mL_{{\rm R},k}(\theta)$.
Then, each local computer should report this local estimator to the central computer.
Next, the central computer assembles those estimators to form a more powerful estimator.
To achieve this goal, a typical assembling
solution is the one-shot type strategy \citep{zhang2013communication,chang2017divide}.
More specifically, the final estimator is given by $\wh\theta_{\rm RMLE}=\sum_{k=1}^{K} \wh\theta_{{\rm RMLE},k}/K$.
The asymptotic distribution of $\wh\theta_{\rm RMLE}$ is presented in the following theorem.

\bet
\label{Theorem 1}
{\color{black} 
Assume (C1) $P(\|Z_i\|>M) \le 2\exp (-C_{\rm Tail} M^2)$ with $Z_i=(1,X_i^\top)^\top \in \mR^{p+1}$ for some positive constant $C_{\rm Tail}$,
(C2) $n\to\infty$ as $N\to\infty$,
(C3) $\log^2 N/(ne^{\alpha_N^*}) = O(1)$, and (\ref{eq: condition}).
Then we have the following asymptotic representation 
\beq
\sqrt{Ne^{\alpha_N^*}} \big(\wh{\theta}_{\rm RMLE}-\theta^*\big) =
P_1 + P_2/2 + o_p\big(K/\sqrt{Ne^{\alpha_N^*}}\big),\nonumber
\eeq
where $P_1 = -(Ne^{\alpha_N^*})^{1/2} K^{-1} \sum_{k=1}^K \ddot{\mL}_{{\rm R},k}^{-1}\big(\theta^*\big) \dot{\mL}_{{\rm R},k} \big( \theta^* \big)$ and 
$P_2 = (Ne^{\alpha_N^*})^{-1/2} K B(\theta^*) $.
Here $B(\theta^*)$ is a random bias term such that $C_{\min} \le E\{\|B(\theta^*)\|\} \le C_{\max}$ for some fixed positive constants $0<C_{\min}<C_{\max}<\infty$.
}
\eet
\noindent

{\color{black}
The theorem condition (C1) requires that covariate distributions have exponentially decayed tail probability \citep{zhang2020concentration}.
The theorem condition (C2) implies that the expected number of the instances on each local computer $n = E(n_k)$ should diverge to infinity as the total sample size $N\to\infty$.
In the meanwhile, we have $E(n_{1k}) = E(\sum_{i=1}^N a_i^{(k)} Y_i) = ne^{\alpha_N^*} E\{e^{X_i^\top\beta^*}/(1+e^{Z_i^\top\theta^*})\}$.
By condition (C3), we require that the number of the positive instances on the local computer should be large enough. 
Then by Theorem \ref{Theorem 1}, we know that $\sqrt{Ne^{\alpha_N^*}}\big(\wh{\theta}_{\rm RMLE}-\theta^*\big)$ can be decomposed into three parts.
The first part is $P_1$, where $P_1 \to_d N(0,\Sigma^{*-1})$ as $N \to \infty$.
The second part $P_2$ is a random bias term being of the order $K/(Ne^{\alpha_N^*})$,
where the analytical formula for $B(\theta^*)$ is given in Appendix A.1.
The third part is a higher order and negligible term as compared with $P_2$.
If $K$ is sufficiently small in the sense of $K/\sqrt{Ne^{\alpha_N^*}} \to 0$ as $N\to \infty$, we should have $P_1$ being the leading term.
In this case, $\wh{\theta}_{\rm RMLE}$ shares the same asymptotic distribution as the $\wh{\theta}_{\rm GMLE}$ of \cite{wang2020logistic}.
Otherwise, we should have $P_2/2$ as the dominating term.
This makes the statistical efficiency of $\wh{\theta}_{\rm RMLE}$ poor.
}

\csubsection{Under-Sampling with {\color{black} Unweighted Objective Function} }

Next, we study the asymptotic properties of the distributed estimators by under-sampling.
We start with $\wh\theta_{\rm US}$ utilized by the unweighted loss function \eqref{eq: unweighted likelihood}.
To obtain the US estimator, we distribute the full data $\mS_F$ to each local computer by the COPY strategy.
Let $\mS_k^{C}$ be the sample distributed to the $k$th local computer under the COPY strategy.
Denote $\mS_k^C = \mS_{k+}^C\cup \mS_{k-}^C$, where
$\mS_{k+}^C$ and $\mS_{k-}^C$ refer to the positive and negative instances on the $k$th local computer, respectively.
For the COPY strategy, we have $\mS_{k+}^C = \mS_+$ for $1\le k\le K$, which implies that the positive instances remain the same for all local computers.
As one can see, the advantage of the COPY strategy is that the number of positive cases allocated to each local computer becomes much larger than that of the RANDOM method.
The negative instances are then randomly distributed on each local computer such that
$\cup_k \mS_{k-}^C = \mS_-$ with $\mS_{k_1-}^C\cap\mS_{k_2-}^C = \emptyset$ for any $k_1\ne k_2$.
Let $n_{1k}^C = |\mS_{k+}^C|$ and $n_{0k}^C = |\mS_{k-}^C|$.
We typically require that $n_{1k}^C = O_p(n_{0k}^C)$. In other words, the number of negative instances assigned to each local computer should not be much smaller than that of the positive instances, which is also the most common case in practice.

Subsequently, define a local MLE for each local computer as $\wh\theta_{{\rm US},k}=\argmax_{\theta} \mL_{{\rm US},k}(\theta)$, where we have
\beq
\mL_{{\rm US},k}(\theta)=
\sum_{i=1}^N \Big\{Y_i \log p_i\big(\alpha_N,\beta\big) + \big(1-Y_i\big)a_i^{(k)}\log \big(1-p_i(\alpha_N,\beta) \big)\Big\}.\nonumber
\eeq
and recall that $a_i^{(k)} = 1$ if the $i$th instance is allocated to the $k$th local computer.
As a consequence, the $\wh\theta_{{\rm US},k}$ on each worker is equivalent to the under-sampled estimator proposed by \cite{wang2020logistic}.
After conducting local estimation, each local computer sends the local estimator $\wh \theta_{{\rm US},k}$ to the central computer.
Similarly, by using the one-shot strategy, we obtain the final estimator as
$\wh\theta_{\rm US}=\sum_{k=1}^{K} \wh\theta_{{\rm US},k}/K$.
We next analyze the asymptotic properties of $\wh\theta_{\rm US}$ in the following theorem.

\bet
\label{Theorem 2}
{\color{black} Assume the same conditions in Theorem \ref{Theorem 1}.} 
Define $\flat=(\log K,0,\cdots,0)$ and $\Sigma^*_2 = E\big\{ \big(1+\gamma e^{X_i^\top \beta^*}\big)^{-1} e^{X_i^\top \beta^*}Z_iZ_i^\top \big\}$ with $\gamma = \lim_{N\to\infty} K e^{\alpha_N^*} \in [0,\infty)$.
We then have the following asymptotic representation as
\beqrs
\sqrt{Ne^{\alpha_N^*}}\big(\wh{\theta}_{\mathrm{US}}- \theta^* - \flat \big)=
\big(Ne^{\alpha_N^*}\big)^{-1/2} \Sigma^{*-1}_2 K^{-1} \sum_{k=1}^{K} \dot{\mL}_{{\rm US},k} \big( \theta^*+\flat \big) + {\color{black} o_p (1)}.
\eeqrs
\eet
\noindent

By Theorem \ref{Theorem 2}, we know that $\sqrt{Ne^{\alpha_N^*}}\big(\wh{\theta}_{\mathrm{US}}- \theta^* - \flat \big)$ can be decomposed into two parts.
For the first part, we have $\big(Ne^{\alpha_N^*}\big)^{-1/2} \Sigma^{*-1}_2
K^{-1}\sum_{k=1}^{K} \dot{\mL}_{{\rm US},k} \big( \theta^*+\flat \big) \to_d N\big( 0,\Sigma^{*-1}_2\Sigma^*_1\Sigma^{*-1}_2 \big)$ as $N \to \infty$, where $\Sigma^*_1 = E\big\{ ( 1+\gamma e^{X_i^\top \beta^*} )^{-2}
e^{X_i^\top \beta^*} Z_iZ_i^\top \big\}$.
{\color{black} The second part is a higher order negligible term.}
{\color{black} Here the asymptotic normality can be established since $\big(Ne^{\alpha_N^*}\big)^{-1/2} \Sigma^{*-1}_2 K^{-1} \sum_{k=1}^{K} \dot{\mL}_{{\rm US},k} \big( \theta^*+\flat \big)$ can be written as the summation of a set of carefully defined independent random variables; see Theorem 2 Step 4 in Appendix A.2 for details.
Therefore, the Lindeberg-Feller Central Limit Theorem can be readily applied.}
Consequently, $\wt{\theta}_{\rm US}=\wh \theta_{\rm US} - \flat$ is $\sqrt{Ne^{\alpha_N^*}}$-consistent for $\theta^*$.
Comparing this results with that of Theorem \ref{Theorem 1}, we find some interesting differences.
First, an additional bias correction term $\flat$ is necessarily involved for the intercept. It is mainly caused by the distortion of the data distribution in the US setting.
Second, we find that
the US estimator has a lower bias than that of the RMLE estimator if $K$ is large.
That is mainly because the bias of the local estimators computed by the COPY strategy is smaller than that of the RANDOM strategy.

We further comment about the constant $\gamma$ occurring in both $\Sigma_1^*$ and $\Sigma_2^*$. As remarked by \cite{wang2020logistic}, one can verify that
$\gamma E(e^{X_i^\top\beta^*})\approx N_1/(N_0/K)$ asymptotically, where $N_0/K$ represents the number of negative instances on each local computer.
Thus, $\gamma E(e^{X_i^\top\beta^*})$  asymptotically quantifies the ratio of the positive instance number to negative instance number.
If $\gamma = 0$, then the number of negative instances dominates the positive ones.
Therefore, we have $\Sigma^{*-1}_2\Sigma^*_1\Sigma^{*-1}_2 = \Sigma^{*-1}$.
This implies that the US estimator shares the same asymptotic covariance matrix as the GMLE $\wh\theta_{\rm GMLE}$.
If $0<\gamma<\infty$, then the positive and negative instances are of comparable sizes.
This implies that the US estimator becomes statistically inefficient as compared with the GMLE $\wh\theta_{\rm GMLE}$.
This finding is also consistent with Theorem 2 in \citep{wang2020logistic}.
We do not consider $\gamma= \infty$, which implies that the number of positive instances is much larger than that of the negative ones.

\csubsection{Under-Sampling with Weighted {\color{black} Objective Function} }

The analysis presented in Sections 3.1 and 3.2 suggests that neither the RMLE nor the US estimator can achieve the global asymptotic efficiency.
The RMLE fails because too small amount of positive instances are distributed to each local computer.
The US estimator fails since the {\color{black} US objective function} used by each local computer is not unbiased for the global one.
We are then inspired to develop a new local log-likelihood function, which should be an unbiased estimator for the global one.
Meanwhile, all positive instances should be used by each local machine.
To this end, we propose an IPW estimator as follows.
Specifically, we still distribute the full data $\mS_F$ to each local computer by the COPY strategy.
Next, we define for each local computer a local MLE as $\wh\theta_{{\rm IPW},k}=\argmax_{\theta}\mL_{{\rm IPW},k}(\theta)$, where we have
\beq
\mL_{{\rm IPW},k}(\theta)=
\sum_{i=1}^N \Big\{Y_i \log p_i\big(\alpha_N,\beta\big) + K \big(1-Y_i\big) a_i^{(k)}\log \big(1-p_i(\alpha_N,\beta) \big) \Big\}.\nonumber
\eeq
Hence, on each worker, the $\wh\theta_{{\rm IPW},k}$ can be treated as the under-sampled weighted estimator proposed by \cite{wang2020logistic} as also given in (\ref{eq: weighted likelihood}).
One can immediately verify that $E\{\mL_{{\rm IPW},k}(\theta)|\mS_F\} = \mL \big(\theta\big)$, where recall that $\mS_F=\big\{\big(X_{i}, Y_i\big): 1\le i\le N \big\}$ denotes the full data.
Then, each local computer sends this local estimator $\wh\theta_{{\rm IPW},k}$ to the central computer.
Similarly, by using the one-shot strategy, we obtain the
final estimator as
$\wh\theta_{\rm IPW}=\sum_{k=1}^{K} \wh\theta_{{\rm IPW},k}/K$.
As noted before, $\mL_{{\rm IPW},k}(\theta)$ is now an unbiased estimator for the global log-likelihood function, and we expect
$\wh\theta_{\rm IPW}$ to achieve the same asymptotic efficiency as the GMLE.
To this end, we analyze the asymptotic properties of $\wh\theta_{\rm IPW}$ in the following theorem.

\bet
\label{Theorem 3}
{\color{black} Assume the same conditions in Theorem \ref{Theorem 1},} we then have the following asymptotic representation as
\beqrs
\sqrt{Ne^{\alpha_N^*}}\big(\wh{\theta}_{\mathrm{IPW}}-\theta^*\big)=
\big(Ne^{\alpha_N^*}\big)^{-1/2} \Sigma^{*-1} \dot{\mL} \big( \theta^* \big)
+ {\color{black}o_p (1)}.
\eeqrs
\eet
\noindent

By Theorem \ref{Theorem 3}, we know that $\sqrt{Ne^{\alpha_N^*}}\big(\wh{\theta}_{\mathrm{IPW}}-\theta^*\big)$ could be decomposed into two parts.
For the first part, we have $\big(Ne^{\alpha_N^*}\big)^{-1/2} \Sigma^{*-1} \dot{\mL} \big( \theta^* \big) \to_d N\big(0,\Sigma^{*-1}\big)$ as $N \to \infty$.
{\color{black} The second part is of the order $o_p(1)$, which is a higher order negligible term.}
Consequently, $\wh{\theta}_{\rm IPW}$ is $\sqrt{Ne^{\alpha_N^*}}$-consistent for $\theta^*$.
Comparing this result of the GMLE in \cite{wang2020logistic}, we find that $\wh\theta_{\rm IPW}$ shares the same asymptotic distribution as the GMLE.
{\color{black} Comparing the result of the US estimator in Theorem \ref{Theorem 2},
we find that the US estimator over-weights the positive instances by using the US objective function (2.4) on the local computers. 
However, the IPW estimator assigns equal weights to positive instances and negative instances by using the IPW objective function (2.5) on the local computers.
This is the key reason why the IPW estimator performs better than the US estimator.}
Particularly, the $\gamma$ given in Theorem \ref{Theorem 2} is not involved, which represents the asymptotic ratio of positive instances to negative instances.
As a consequence, we do not require
$\gamma = 0$ to attain the global efficiency as compared to the US estimator
(or the under-sampled estimator in \cite{wang2020logistic}).
Our extensive numerical studies also illustrate better finite sample performance of $\wh\theta_{\rm IPW}$.
To summarize, the RMLE estimator $\wh\theta_{\rm RMLE}$ with a large $K$ suffers from significant bias.
The debiased US estimator $\widetilde{\theta}_{\rm US}$ is statistically inefficient either due to its high asymptotic covariance
if the number of negative instances distributed on each computer node is not enough.
The IPW estimator $\wh\theta_{\rm IPW}$ stands out as the most attractive estimator.

It is remarkable that both the US and IPW estimators investigated in \cite{wang2020logistic} are different from their counterpart estimators studied in our work.
Specifically, these two estimators in \cite{wang2020logistic} are based on a subsample, which contains all positive instances but only a small fraction of negative instances.
By doing so, a significant amount of computation cost can be nicely saved.
In this case, \cite{wang2020logistic} found that the US estimator is more efficient than the IPW estimator.
However, both the US and IPW estimators studied in our work are based on the whole sample but computed in a distributed way.
Therefore, for our estimators, not only all positive instances but also all negative instances are fully used.
In fact, all the positive instances are repeatedly used by different local computers due to our COPY strategy.
In contrast, only a small proportion of negative instances are used in \cite{wang2020logistic}.
This makes the theoretical properties of our US and IPW estimators very different from those of \cite{wang2020logistic}.
This is also the key reason accounting for the performance differences between the two sets of estimators.

\csection{NUMERICAL STUDIES}

\csubsection{A Simulation Study}

\begin{center}
\textit{Model Setup and Performance Measure}
\end{center}

To demonstrate the finite sample performance of the proposed methods, a number of simulation studies are conducted in this section.
A standard logistic regression model \eqref{eq: logistic} is used to generate the full data with covariate $Z_i=(1,X_i^\top)^\top \in \mR^5$.
Here the covariates $X_i$s are generated from $N(0,\Sigma)$ with $\Sigma=(\sigma_{ij})$ and $\sigma_{ij} = 0.2^{|i-j|}$.
The total sample sizes are $N=10^4,10^5, 5\times10^5$ and $10^6$.
For a fixed $N$, we set $\alpha_N^*=-0.45\log N$ and $\beta^*=(1,1,1,1)^\top$.
By doing so, we allow
$P(Y=1) \to 0$ and $E\big(N_{1}\big) \to \infty$ as $N \to \infty$.
We next set the number of local computers (i.e., $K$) in two different cases.
For CASE 1, we set $K=17, 36, 63, 81$ with the four different sample sizes, respectively.
One can verify that, for the COPY strategy, the number of positive instances is approximately 1.5 times as large as that of the negative instances on each local computer.
In contrast, for CASE 2, we set $K=2,3,4,5$ accordingly.
By doing so, the number of negative instances assigned to each local computer should be much larger than that of positive instances for the COPY strategy.
Next, two different distribution strategies (i.e. RANDOM and COPY) are considered.
We then obtain three local estimators $\wh\theta_{{\rm RMLE},k}$, $\widetilde\theta_{{\rm US},k}$, and  $\wh\theta_{{\rm IPW},k}$ for every local machine $k$.
Here $\widetilde\theta_{{\rm US},k}$ and $\wh\theta_{{\rm IPW},k}$ can be treated as the under-sampled estimators proposed by \citep{wang2020logistic}.
This leads to the combined estimators as $\wh\theta_{{\rm RMLE}}$, $\widetilde\theta_{{\rm US}}$, and $\wh\theta_{\rm IPW}$ on the central computer.
Here for the US method, we use the debiased estimator $\wt\theta_{{\rm US}}$ (instead of $\wh\theta_{{\rm US}}$) as our final estimator.
For comparison purpose, the GMLE $\wh\theta_{\rm GMLE}$ is also calculated.
For a reliable evaluation, each experiment is randomly replicated for a total of $M = 500$ times.
Let $\wh\theta^{(m)} =(\theta_j^{(m)}:1\le j\le p+1)^\top$ be one particular estimator obtained in the $m$-th replication (e.g., $\wh\theta_{{\rm RMLE},k}$ for $k=1$ or $\wh\theta_{\rm RMLE}$).
To evaluate the estimation accuracy, we calculate the Root Mean Square Error (RMSE) as $\mbox{RMSE}=(p+1)^{-1}\sum_{j=1}^{p+1} \big\{M^{-1} \sum_{m=1}^{M}(\wh{\theta}_j^{(m)} -\theta_j^*)^2\big\}^{1/2}$.
Then the RMSE of $\wh\theta_{\rm GMLE}$ is numerically computed according to its theoretical formula.
Furthermore, the absolute bias of $\wh\theta$ is estimated by BIAS$=(p+1)^{-1}\sum_{j=1}^{p+1} |\bar\theta_j-\theta_j^* |$, where $\bar\theta_j = M^{-1}\sum_{m=1}^M \wh\theta_j^{(m)}$.
The standard error (SE) of $\wh\theta$ is estimated by $\mbox{SE} = (p+1)^{-1}\sum_{j=1}^{p+1} \{M^{-1}
\sum_{m=1}^M (\wh\theta_j^{(m)}-\bar\theta_j)^2\}^{1/2}$.

\begin{center}
\textit{Simulation Results}
\end{center}

\begin{figure}[htb]
\centering
\includegraphics[width=0.92 \textwidth]{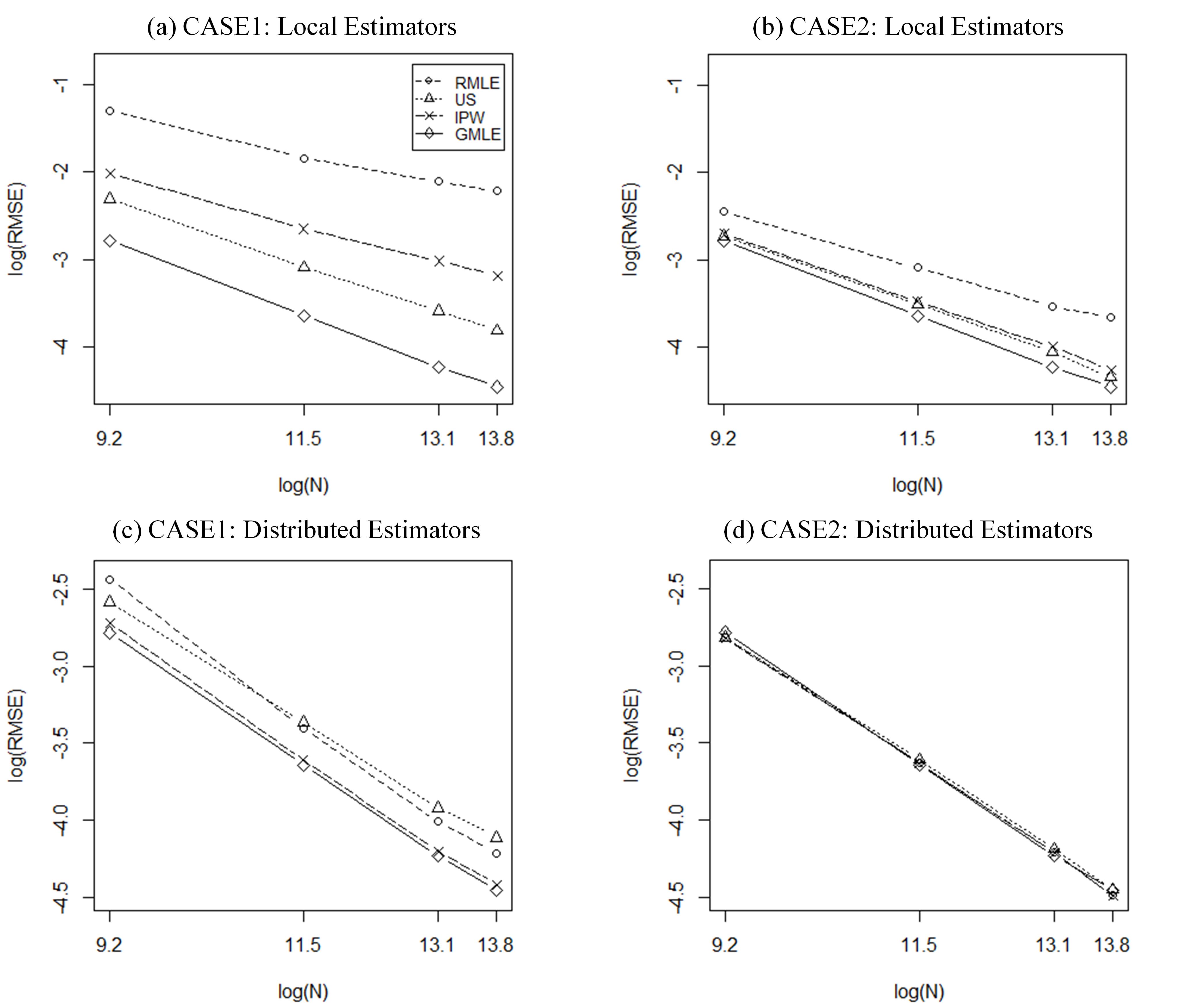}\hfill
\caption{RMSE of the local and distributed estimators in log-scale.
The horizontal axis presents the total sample size $N$ in log-scale.
The top panels show the local estimators.
The bottom panels present the distributed estimators.
The left panels show the cases where the number of positive cases is approximately 1.5 times as large as that of negative ones.
The right panels present the cases where the number of negative instances is much larger than that of positive ones.
}
\label{f: RMSE}
\end{figure}

The detailed results are given in Figure \ref{f: RMSE}.
Here we study both the local and distributed estimators.
Two different cases (i.e., CASE 1 and CASE 2) regarding the number of local machines are considered.
This leads to a total of four combinations that are represented in different panels.
The vertical axis in Figure \ref{f: RMSE} represents the RMSE value in log-scale.
The horizontal axis denotes the total sample size also in log-scale.
First, the top left panel presents the results of the local estimators for CASE 1.
In this case, all estimators under study are much less efficient than the GMLE in the sense that the $\log$(RMSE) values of various estimators are much larger than that of the GMLE.
This is because other estimators (i.e., $\wh\theta_{{\rm RMLE},k}$, $\widetilde\theta_{{\rm US},k}$ and $\wh\theta_{{\rm IPW},k}$) are local estimators.
Here $\wt\theta_{{\rm US},k} = \wh\theta_{{\rm US},k}-\flat$ is the debiased estimator with $\flat=(\log K,0,\cdots,0)$.
The sample sizes used by these estimators are much smaller than that of the global estimator.
Consequently, they are expected to be less efficient than the global estimator.
However, among all local estimators, we find that the performance of $\wh\theta_{{\rm RMLE},k}$ is always the worst. This is expected because the number of positive instances used by the RMLE estimator is much less than that of other local estimators.
Comparatively speaking, we find that $\widetilde\theta_{{\rm US},k}$ performs better than $\wh\theta_{{\rm IPW},k}$.
These observations are in line with that of \cite{wang2020logistic}.

The top right panel in Figure \ref{f: RMSE} presents the results of the local estimators for CASE 2.
Compared with the top left panel, we find that the GMLE remains to be the best estimator.
However, among all local estimators, the performance differences are markedly smaller.
This is because the number of negative instances assigned to the local machine is sufficiently large in this case.
This makes the performances of all local estimators improve towards that of the global estimator and their relative differences vanish.

The bottom left panel presents the $\log$(RMSE) values of distributed estimators for CASE 1.
We find that the performances of the distributed estimators (e.g., $\wh\theta_{\rm IPW}$) are improved compared to the local estimators (e.g., $\widetilde\theta_{\rm US,k}$ and $\wh\theta_{\rm IPW, k}$ proposed by \citep{wang2020logistic}) especially when the under-sampled negative instances are not enough.
For example, the RMSE value of $\wt\theta_{\rm US,k}$ is 0.094 and the RMSE value of $\wh\theta_{\rm IPW,k}$ is 0.121 when $N=10^4$ in the top left panel.
For comparison, the RMSE value of $\wh\theta_{\rm IPW}$ is 0.065, which is close to that of the GMLE (i.e., 0.061).
This implies that the IPW estimator is less sensitive to the ratio of positive to negative instances.
Among all distributed estimators, we find that the debiased US estimator $\wt\theta_{\rm US}$ appears to be the worst estimator in the sense that the associated $\log$(RMSE) value is always the largest.
In contrast, the IPW estimator $\wh\theta_{\rm IPW}$ stands out to be the best estimator.
The relative difference among different distributed estimators disappears as the number of negative instances assigned to each local machine increases.
This can be seen from the results of the bottom right panel.
{\color{black} 
More detailed results about Figure \ref{f: RMSE}(c) are given in Table \ref{t: distributed}.
By Table \ref{t: distributed}, we find that the RMLE estimator $\wh\theta_{\rm RMLE}$ in CASE 1 demonstrates a large bias, since $K$ is relatively large.
In the meanwhile, the debiased US estimator $\wt\theta_{\rm US}$ suffers from high SE values.
These observations are in line with the theoretical findings of the proposed Theorems \ref{Theorem 1}--\ref{Theorem 2}.
}

\begin{table}[htb]
\begin{center}
\caption{Simulation Results for the Distributed Estimators under CASE 1.}
\label{t: distributed}
\renewcommand\arraystretch{1.5} 
\resizebox{\hsize}{!}{
\begin{tabular}{ccccccccccccccccccccccccccc}
\hline \hline
&&& \multicolumn{3}{c}{RMLE} && \multicolumn{3}{c}{US} &&
\multicolumn{3}{c}{IPW} \\
\cline{4-6}
\cline{8-10}
\cline{12-14}
$N$ & $\color{black} \sqrt{Ne^{\alpha_N^*}}$ && BIAS & SE & RMSE
&& BIAS & SE & RMSE
&& BIAS & SE & RMSE \\
\hline
$10^4$ & $\color{black} 13$
&& $0.055$ & $0.067$ & $0.088$ && $0.005$ & $0.075$ & $0.076$ && $0.018$ & $0.063$ & $0.066$ \\

$10^5$ & $\color{black} 24$
&& $0.019$ & $0.027$ & $0.033$ && $0.001$ & $0.035$ & $0.035$ && $0.006$ & $0.026$ & $0.027$ \\

$5\times10^5$ & $\color{black} 37$
&& $0.010$ & $0.015$ & $0.018$ && $0.000$ & $0.020$ & $0.020$ && $0.003$ & $0.015$ & $0.015$ \\

$10^6$ & $\color{black} 45$
&& $0.008$ & $0.012$ & $0.015$ && $0.001$ & $0.016$ & $0.016$ && $0.003$ & $0.012$ & $0.012$ \\
\hline
\end{tabular}
}
\end{center}
\end{table}

\csubsection{Sweden Traffic Sign Data Analysis}

\begin{center}
\textit{Data Processing}
\end{center}

For illustration purpose, we present an interesting real data example.
The dataset used in this study is the Sweden Traffic Sign (STS) dataset, which is publicly available at \textit{https://www.cvl.isy.liu.se/research/datasets/traffic-signs-dataset/}.
It contains a total of 1,970 annotated images with various traffic signs annotated by bounding boxes; see Figure \ref{f: bbox} for a graphical illustration.
We aim to detect the traffic signs in Figure \ref{f: bbox} automatically.
For a reliable evaluation, we randomly split the entire data into two parts.
The first part contains 1,576 images (about $80\%$ of the whole data) for training, while the remaining 394 images (about $20\%$ of the whole data) for testing.
This task contains two important steps; see \cite{girshick2014rich,girshick2015fast}.
For the first step, one needs to automatically detect a sufficiently tight local region containing a traffic sign from an input image without bounding box information.
In the second step, one needs to classify the traffic signs detected in the local region to different categories (e.g., prohibitive, informative, warning and mandatory traffic signs).
In this study, we focus on the first step.
We subsequently demonstrate how this task can be converted into a logistic regression problem, which has a large sample size and can be efficiently solved by our proposed method in a distributed way.

\begin{figure}[htbp]
\begin{adjustbox}{addcode={
\begin{minipage}{\width}}{
\caption{
	Illustration of the data preprocessing pipeline for one particular image.
	The top panel illustrates how the nonlinear features are generated.
	The bottom panel shows how the response is generated.
}
\label{f: pipeline}
\end{minipage}},rotate=90,center}
\includegraphics[width=24cm]{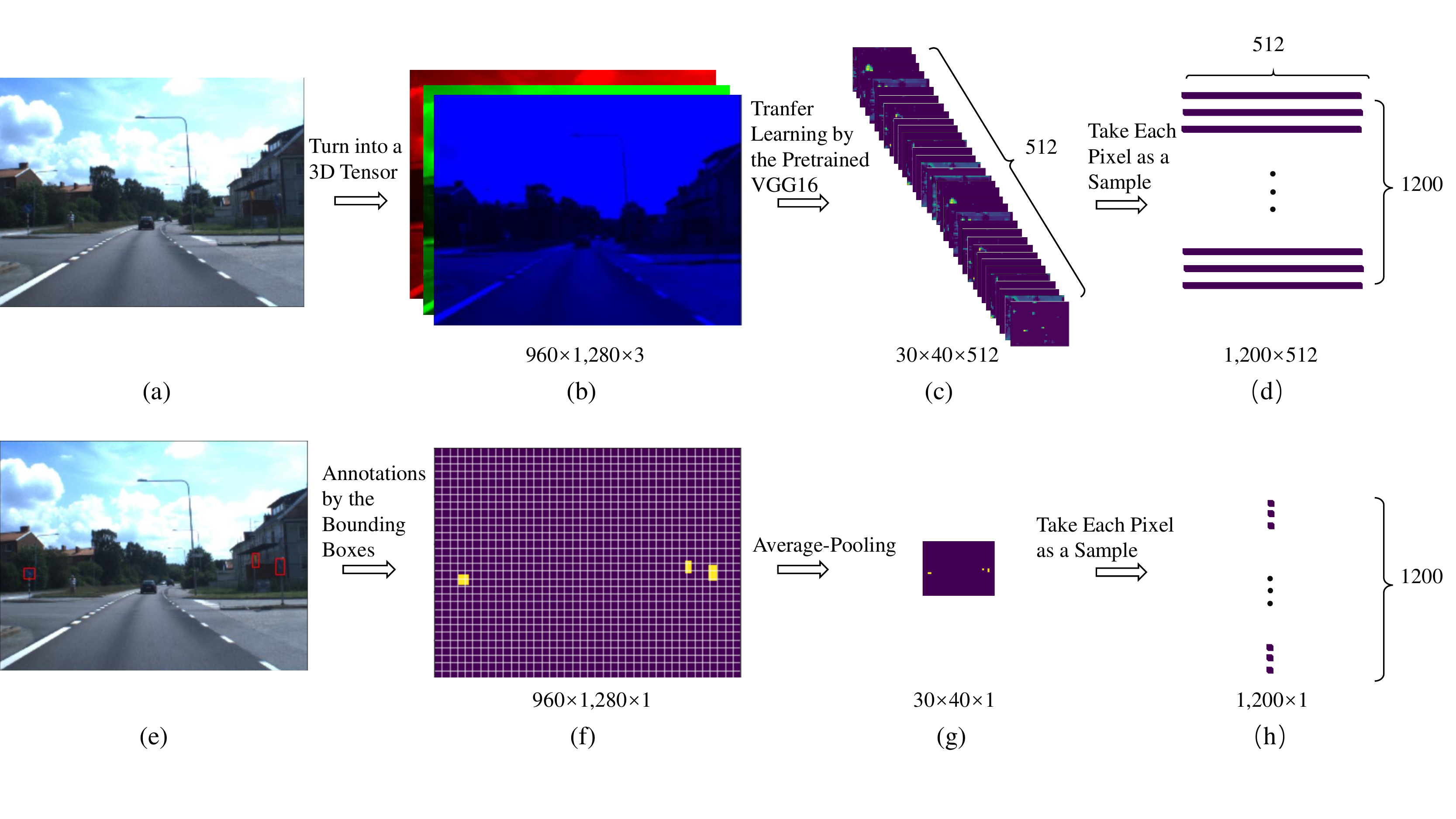}%
\end{adjustbox}
\end{figure}

Specifically, each image given in the STS dataset is of relatively high resolution; see Figure \ref{f: pipeline}(a).
Mathematically, each image can be represented by a tensor of size $960 \times 1,280 \times 3$; see Figure \ref{f: pipeline}(b).
Next, we apply a pretrained VGG16 model on the image \citep{simonyan2014very}.
The VGG16 model is a classical convolutional neural network model with a total of 13 convolutional layers. The last two fully connected layers are dropped.
Then, a feature map of size $30 \times 40 \times 512$ can be extracted from the last convolutional layer; see Figure \ref{f: pipeline}(c).
This can be viewed as a new ``image" of resolution $30 \times 40$ but with a total of 512 channels.
We then treat each pixel of this feature map as one sample.
As a result, a total of $30\times40 = 1, 200$ pixel samples can be generated for every single image.
For each pixel sample, a feature vector of 512 dimension can be constructed.
Consequently, we have $p=512$ in this case.
The $i$th image is then denoted by $\mX_{i,k_1,k_2} \in \mR^{512}$ with $1\le i \le N$, $1\le k_1 \le 30$ and $1\le k_2 \le 40$; see Figure \ref{f: pipeline}(d).
Then the total sample size is given by $N = 1,970\times 1,200 = 2,364,000$.

We next present the details about how the response $\mY_{i,k_1,k_2} \in \{0,1\}$ is constructed.
Define $W_i=(W_{i,k_1,k_2})$ be a binary matrix with dimensions $960 \times 1,280$ and $W_{i,k_1,k_2} \in \{0,1\}$.
For a given image with the bounding box information, define $W_{i,k_1,k_2}=1$ if the $(k_1,k_2)$th pixel is located in the bounding box region and $W_{i,k_1,k_2}=0$ otherwise; see Figure \ref{f: pipeline}(e) and (f).
Subsequently, we partition $W_i$ matrix into a $30\times40$ block matrix with equal sizes; see Figure \ref{f: pipeline}(f).
Specifically, we write this block matrix as $\mW_i =( \mW_{i,k_1,k_2} )$ with $\mW_{i,k_1,k_2} \in \mR^{32 \times 32}$, $1\le k_1 \le 30$ and $1\le k_2 \le 40$.
Next, compute the average value of the block matrix $\mW_{i,k_1,k_2}$ and denote it by $\mu_{i,k_1,k_2}$.
With the help of TensorFlow and GPU, this operation can be efficiently conducted in a fully parallel way by an average pooling operation.
Define $\mY_{i,k_1,k_2} = I(\mu_{i,k_1,k_2}>0.5)$. Then $\mY_{i,k_1,k_2}$ becomes the binary response associated with $\mX_{i,k_1,k_2}$; see Figure \ref{f: pipeline}(g) and (h).
They both correspond to the same region in the original image.
All data ($8.59$ GB, including $\mX_{i,k_1,k_2}$ and $\mY_{i,k_1,k_2}$) are placed on the hard drive.
A simple calculation reveals that the sample mean of $\mY_{i,k_1,k_2}$ is $0.225\%$, which is extremely small. Thus, we can treat it as the rare events data.


Since the total sample size is extremely large, we call for a distributed computation.
For illustration purpose, we fix the number of local computers as $K=50$.
This leads to the sample size allocated to each local machine being approximately $N/K=47,280$ by the RANDOM strategy and $N_1+N_0/K=52,491$ by the COPY strategy.
Consequently, the three distributed estimators $\wh\theta_{\rm RMLE}$, $\widetilde\theta_{\rm US}$ and $\wh\theta_{\rm IPW}$ are computed based on the train data.
For comparison purpose, $\wh\theta_{\rm GMLE}$ is also computed by self-developed Newton-Raphson type algorithm.
If this algorithm is executed on one single computer, then the time cost is extremely high.
If the algorithm is executed on a distributed system, then the communication cost is extremely high due to the Newton-Raphson type iteration.
Simply speaking, this self-developed algorithm is mainly developed here for theoretical comparison.
It is can hardly be used in real practice due to its high cost in time, either due to communication or computation.

\begin{center}
	\textit{Performance Results}
\end{center}

\begin{figure}[htbp]
\centering
\includegraphics[width=0.5\textwidth]
{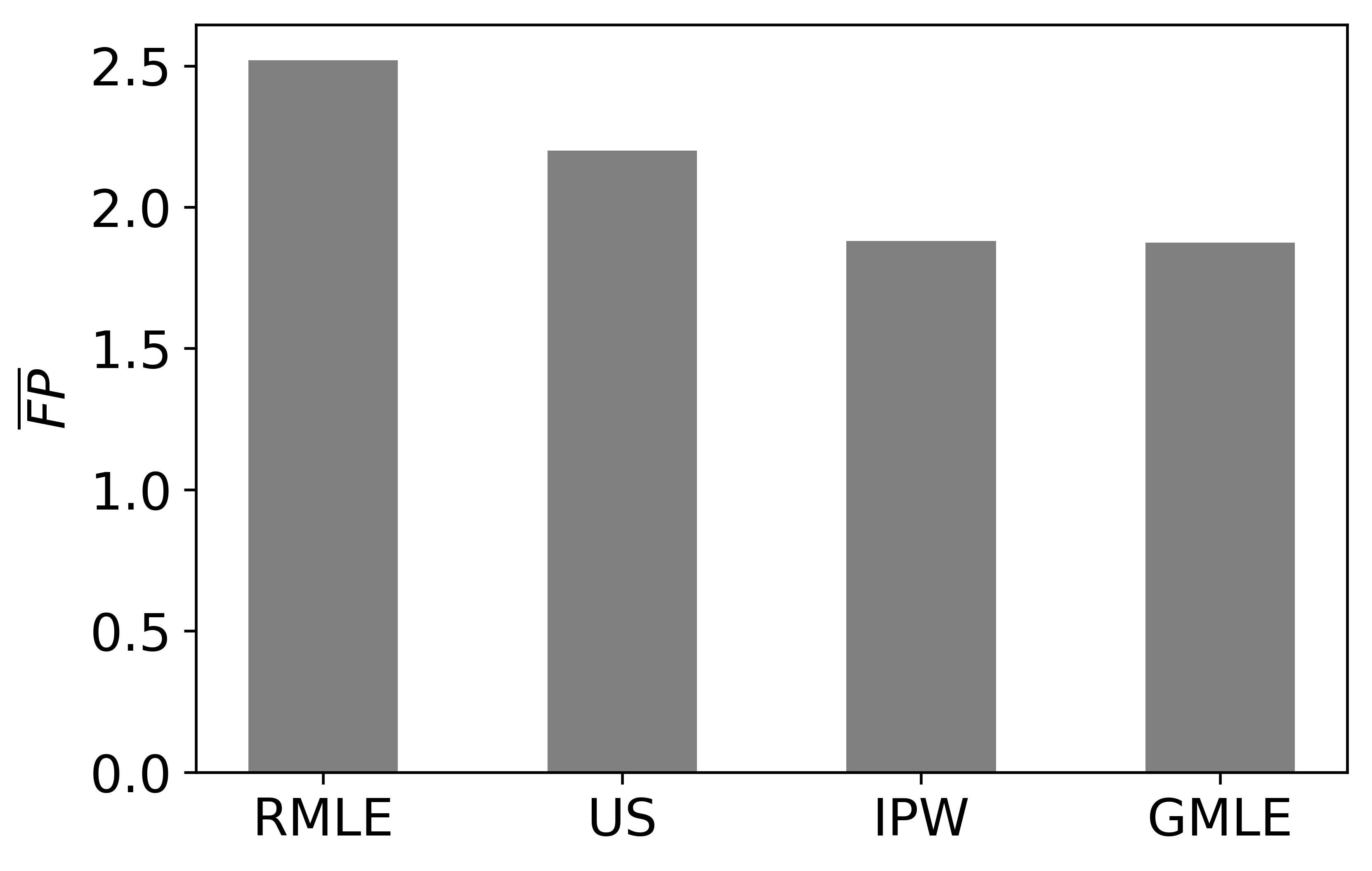}\hfill
\caption{Prediction results $\overline{\rm FP}$ obtained on the test data.}
\label{f: FP}
\end{figure}

Next, consider the $i^*$th image ($1\le i^* \le N^*$) in the test data, where $N^*=394$ denotes the number of images for testing.
For a given pixel $(k_1,k_2)$ in the $i^*$th image and one particular estimator $\wh\theta$ obtained on the train data (i.e., $\wh\theta_{\rm RMLE}$), we then estimate the response probability by $\wh{p}_{i^*,k_1,k_2}=e^{ \wh\theta^{\top} \mX_{i^*,k_1,k_2} }/(1+e^{\wh\theta^{\top} \mX_{i^*,k_1,k_2} })$ and predict $\wh\mY_{i^*,k_1,k_2} = I\big(\wh{p}_{i^*,k_1,k_2}>c_{i^*}\big)$,
where $c_{i^*} = \min \big\{\wh{p}_{i^*,k_1,k_2}: \mY_{i^*,k_1,k_2}=1, 1\le k_1 \le 30,1\le k_2 \le 40 \big\}$.
This $c_i^*$ is the largest threshold value so that all the positive instances can be correctly captured.
However, the price paid here is the false positive predictions.
Define the number of the false positive instances for the $i^*$th image in the test data as ${\rm FP}_{i^*} = \sum_{k_1,k_2} I\big(\wh{\mY}_{i^*,k_1,k_2}=1 \big) I\big(\mY_{i^*,k_1,k_2}=0 \big)$.
Its median value is then computed as ${\rm FP^*}$.
Then its overall mean across different random replications is denoted as $\overline{\rm FP}$.
The prediction results are shown in Figure \ref{f: FP}.
By Figure \ref{f: FP}, we observe that the $\overline{\rm FP}$ value of the IPW method is as low as 1.88, which is much smaller than 2.52 of the RMLE method and 2.20 of the US method.
This value is the same as $1.88$ of the GMLE method.
To summarize, among all distributed estimators, the IPW estimator achieves the best performance with the smallest $\overline{\rm FP}$ value $1.88$.

\begin{figure}[htbp]
\centering
\includegraphics[width=1\textwidth]{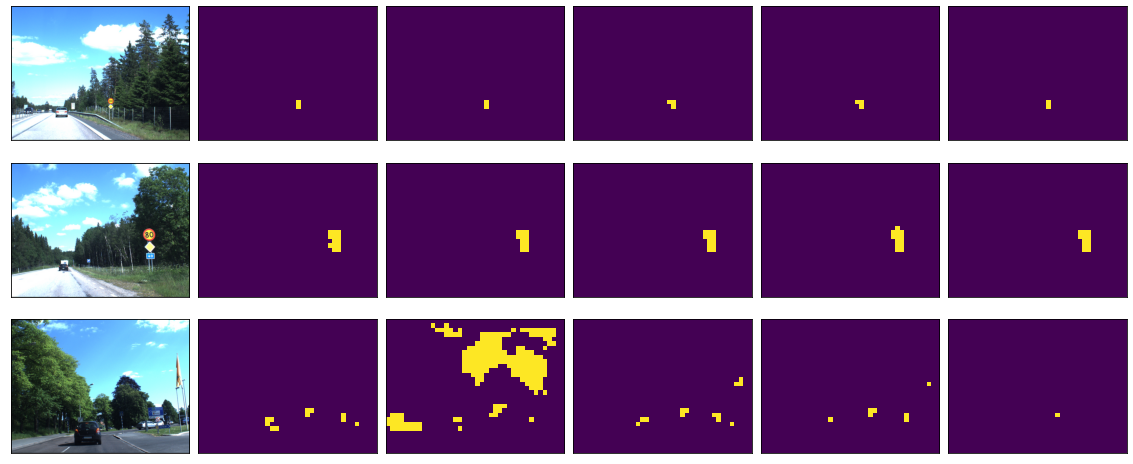}\hfill
\caption{
	A number of arbitrarily selected test examples for prediction demonstration.
	Each row represents one arbitrarily selected image from the test data.
	The input images are shown in the first column.
	The second column presents the US method.
	The third column presents the RMLE method.
	The fourth column presents the IPW method.
	The fifth column presents the GMLE method.
	The last column shows the true annotated regions.
}
\label{f: preds}
\end{figure}

To gain further intuitive understanding about the prediction accuracy, we presents a number of randomly selected prediction results in Figure \ref{f: preds}.
Specifically, each row in Figure \ref{f: preds} shows one arbitrarily selected image in the test data.
The first column shows the original input image of size  $960 \times 1,280$.
The second column presents the prediction results of the US method.
The third column illustrates the prediction results by the RMLE method.
The fourth column presents the prediction results due to the GMLE method.
The fifth column illustrates the prediction results by the IPW method.
The last column represents the true annotated regions.
By Figure \ref{f: preds}, we find that the prediction results of both US and RMLE methods are very noisy.
The prediction results of both GMLE and IPW methods are much better and very comparable.

\csection{CONCLUSION REMARKS}

In this study, we investigate a distributed logistic regression problem for rare events data with massive sizes.
We study here two different data distribution strategies. They are RANDOM and COPY strategies, respectively.
We also investigate three different estimators.
They are $\wh\theta_{\rm RMLE}$, $\widetilde\theta_{\rm US}$ and $\wh\theta_{\rm IPW}$, respectively.
Our results suggest that the COPY strategy together with the modified log-likelihood function for the IPW estimator is the best choice.
The resulting estimator can be statistically as efficient as the global estimator.
To conclude this article, we would like to discuss a number of interesting topics for future study.
First, we focus on the logistic regression model in this paper.
It is interesting to investigate more complicated and general models in future research projects for the rare events data.
{\color{black}
Second, we use the one-shot strategy for the last step in the distributed estimation.
Although this strategy is efficient in terms of communication, it might not be the best choice if the data are non-randomly distributed across different local machines \citep{zhu2021least}.
In this case, various inverse variance weighting (IVW) methods \citep{lin2011aggregated,zhu2021least,yu2022optimal} can be used.
The key idea of IVW is to take the weighted average of local estimators.
The weights are related to the inverse of the Hessian matrices, which are computed by local computers.
How to combine the IVW idea with our COPY strategy for distributed rare events data analysis seems to be an another interesting topic for future study.
}
Lastly, covariates in large datasets typically have high dimensionality. Thus, how to conduct feature selection or screening based on these distributed estimators is worthy of consideration.

\renewcommand \refname{\centerline{REFERENCES}}
\bibliographystyle{asa}
\bibliography{reference}

\end{CJK}
\end{document}